\documentclass[%
 reprint,
superscriptaddress,
bibnotes,
 amsmath,amssymb,
 aps,
]{revtex4-2}

\usepackage{graphicx}
\usepackage{dcolumn}
\usepackage{bm}

\begin{document}

\preprint{APS/123-QED}

\title{Experimental Demonstration of Nanolaser with sub-\textmu A Threshold Current}

\author{Evangelos Dimopoulos}
\affiliation{DTU Electro, Department of Electrical and Photonics Engineering, Technical University of Denmark, Ørsteds Plads 343, Kgs. Lyngby, DK-2800,
Denmark.}

\affiliation{NanoPhoton - Center for Nanophotonics, Technical University of Denmark, Ørsteds Plads 343, Kgs. Lyngby, DK-2800,
Denmark.}

\author{Meng Xiong}
\affiliation{DTU Electro, Department of Electrical and Photonics Engineering, Technical University of Denmark, Ørsteds Plads 343, Kgs. Lyngby, DK-2800,
Denmark.}
\affiliation{NanoPhoton - Center for Nanophotonics, Technical University of Denmark, Ørsteds Plads 343, Kgs. Lyngby, DK-2800,
Denmark.}

\author{Aurimas Sakanas}
\affiliation{DTU Electro, Department of Electrical and Photonics Engineering, Technical University of Denmark, Ørsteds Plads 343, Kgs. Lyngby, DK-2800,
Denmark.}

\author{Andrey Marchevsky}
\affiliation{DTU Electro, Department of Electrical and Photonics Engineering, Technical University of Denmark, Ørsteds Plads 343, Kgs. Lyngby, DK-2800,
Denmark.}

\author{Yi Yu}
\affiliation{DTU Electro, Department of Electrical and Photonics Engineering, Technical University of Denmark, Ørsteds Plads 343, Kgs. Lyngby, DK-2800,
Denmark.}
\affiliation{NanoPhoton - Center for Nanophotonics, Technical University of Denmark, Ørsteds Plads 343, Kgs. Lyngby, DK-2800,
Denmark.}

\author{Elizaveta Semenova}
\affiliation{DTU Electro, Department of Electrical and Photonics Engineering, Technical University of Denmark, Ørsteds Plads 343, Kgs. Lyngby, DK-2800,
Denmark.}
\affiliation{NanoPhoton - Center for Nanophotonics, Technical University of Denmark, Ørsteds Plads 343, Kgs. Lyngby, DK-2800,
Denmark.}

\author{Jesper Mørk}%

\affiliation{DTU Electro, Department of Electrical and Photonics Engineering, Technical University of Denmark, Ørsteds Plads 343, Kgs. Lyngby, DK-2800,
Denmark.}
\affiliation{NanoPhoton - Center for Nanophotonics, Technical University of Denmark, Ørsteds Plads 343, Kgs. Lyngby, DK-2800,
Denmark.}

 \author{Kresten Yvind}%
 \email{kryv@dtu.dk}
\affiliation{DTU Electro, Department of Electrical and Photonics Engineering, Technical University of Denmark, Ørsteds Plads 343, Kgs. Lyngby, DK-2800,
Denmark.}
\affiliation{NanoPhoton - Center for Nanophotonics, Technical University of Denmark, Ørsteds Plads 343, Kgs. Lyngby, DK-2800,
Denmark.}

\date{\today}

\begin{abstract}
  We demonstrate a photonic crystal nanolaser exhibiting an ultra-low threshold of 730 nA at telecom wavelengths. The laser can be directly modulated at 3 GHz at an energy cost of 1 fJ/bit. This is the lowest threshold reported for any laser operating at room temperature and facilitates low-energy on-chip links.
\end{abstract}

\maketitle


\section{Introduction}
Semiconductor lasers with high efficiency and ultra-low operating energy are required for on-chip optical interconnects to replace the currently used electrical wiring that is limited by excessive power consumption and heat dissipation \cite{Miller2017}. Diode lasers exhibit efficient high-speed operation when operated close to the laser threshold. Therefore, realizing ultra-compact lasers with low threshold current and low drive voltage is an important goal. High refractive index contrast structures like photonic crystals (PhCs) are considered the most promising candidates for on-chip applications since PhC cavities provide high quality-factors (Q-factor) at wavelength-scale mode volumes enabling the required miniaturization of the active material and allow electrical injection in a planar geometry.

In the last decade, electrically-driven PhC lasers have been realized \cite{PhC3_Takeda2013, Takeda2014, Crosnier2017}, and the lowest attainable threshold for PhC lasers on the InP and InP-on-Si platforms has converged to around 5 \textmu A \cite{PhC4_Jeong2013, PhC5_Kuramochi2018} and 10 \textmu A \cite{Takeda2021, Dimopoulos2022} respectively. However, numerical calculations using reasonable laser parameter values \cite{Coldren1995DiodeLA} show that the currently achieved thresholds are still an order of magnitude higher than what is theoretically possible. 

In this work, we experimentally demonstrate a continuous-wave PhC laser with an ultra-low threshold of 730 nA emitting at telecom wavelengths. This is the lowest reported threshold current for any laser at room temperature to date. Furthermore, we achieve operating voltages below 1 V and direct modulation at 3 GHz with an energy cost of 1 fJ/bit. This breakthrough is achieved by a laser design that reduces the doping-induced losses, allowing the down-scaling of the active region.
\begin{figure}[!b]
    \centering
    \includegraphics[width=1\linewidth]{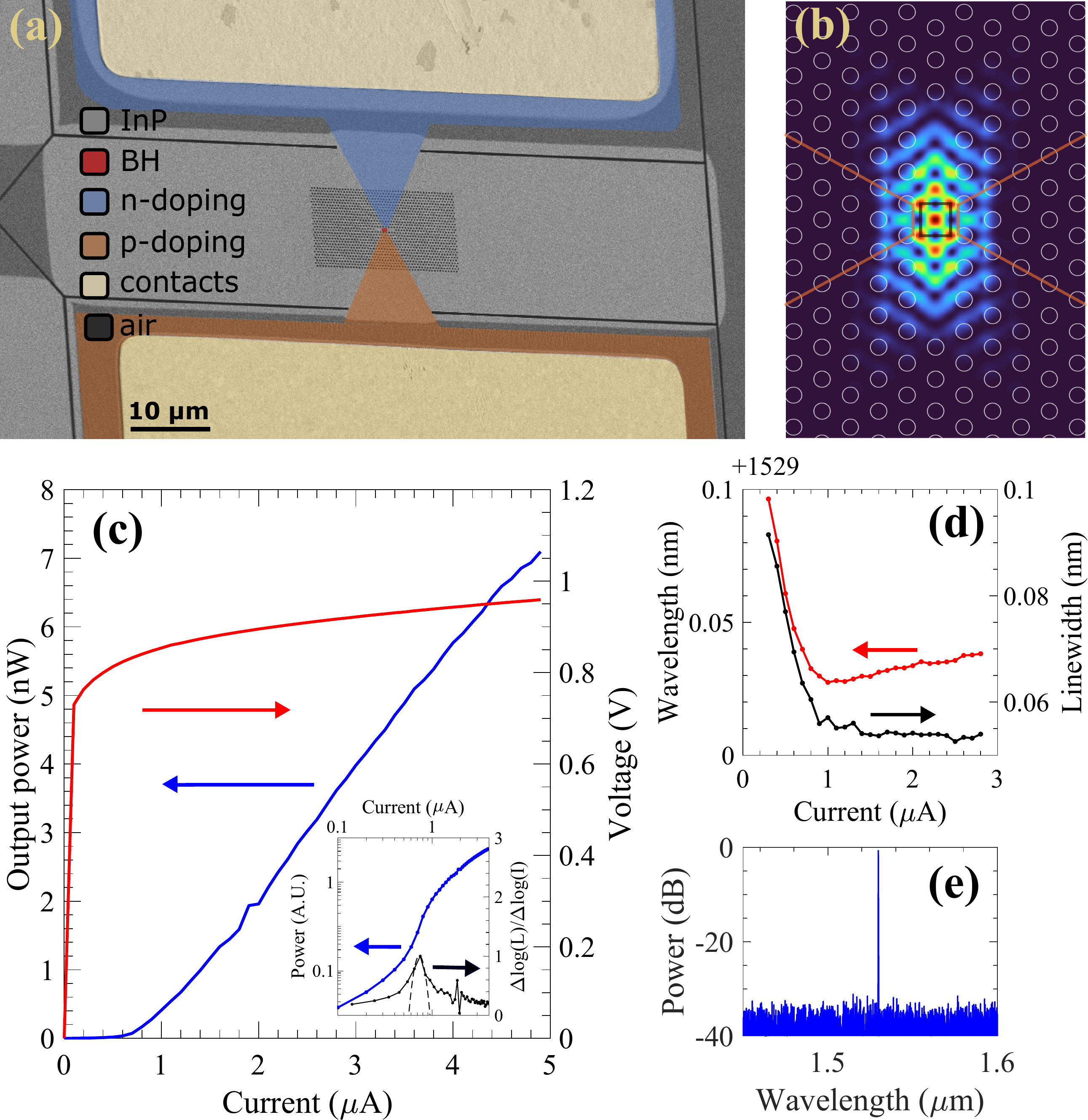}
    \caption{(a) Colored SEM image of the fabricated device. (b) Simulated electric-field profile; the PhC holes and the active and doping regions are outlined in white, black, and orange, respectively. (c) Collected output power and voltage versus input current. Inset: Log-Log plot of the LI curve and its derivative. (d) Laser wavelength and linewidth as a function of current. (e) Laser spectrum at 5 \textmu A.}
    \label{fig:opt1}
\end{figure}
\section{Device Overview}
A scanning electron micrograph (SEM) of the device is shown in Fig. \ref{fig:opt1}(a). The laser is based on a 250 nm InP PhC membrane heterogeneously bonded on a SiO\textsubscript{2}/Si platform via direct bonding \cite{Sakanas2019}. A PhC cavity with 7 missing air holes (L7 cavity) is used to confine the photons \cite{Saldutti2021}, and a buried-heterostructure (BH) active region is used to confine the carriers vertically and laterally. The PhC lattice constant, \textit{a\textsubscript{PhC}}, and hole radius were chosen as 440 nm and 120 nm, respectively, to achieve resonance at telecom wavelengths. The BH active material is based on a single InGaAsP/InAlGaAs-based quantum well of 440x400x8 nm\textsuperscript{3} volume, and its emission is centered around 1550 nm. A lateral carrier injection scheme is used to inject carriers in the BH region. The n-doping and p-doping regions have been formed via Si ion implantation and Zn thermal diffusion, respectively.

The electric field profile of the lasing mode is shown in Fig. \ref{fig:opt1}(b). Reducing the size of the active material is generally beneficial for low power consumption; however, a high Q-factor is critical due to the corresponding decrease in the modal gain. In previous work, we showed that the optical absorption of the p-doping region is the main loss channel that ultimately determines the Q-factor of the laser cavity \cite{Dimopoulos2022}. Since the doping profile normally matches the BH length, we opted for a longer optical cavity to reduce the mode's overlap with the doping region, accessing sub-\textmu A thresholds. 



\section{Results}
In Fig. \ref{fig:opt1}(c), the collected optical power and the applied voltage versus the injection current are shown. Part of the laser light that leaks vertically in the free space is collected via a 50x long-working distance objective coupled to a multi-mode fiber and is then measured using an optical spectrum analyzer (OSA). The threshold current is 730 nA and was extracted from the double-log plot of the L-I curve shown in the inset of Fig. \ref{fig:opt1}(c). Due to a high $\beta$-factor, the threshold was calculated numerically as the maximum of the derivative of the log$(L)$-log$(I)$ curve.

The evolution of the laser wavelength and the linewidth close to the threshold is shown in Fig. \ref{fig:opt1}(d). At the spontaneous emission regime, the resonance blueshifts due to the band-filling effect. Above the threshold, the carrier density is clamped, and the lasing wavelength redshifts due to thermal effects. The linewidth exhibits a kink near the threshold as its value saturates at the resolution limit of the OSA. The laser achieved single-mode operation, as shown in the emission spectrum displayed in Fig. \ref{fig:opt1}(e).

\begin{figure}[t]
    \centering
    \includegraphics[width=0.92\linewidth]{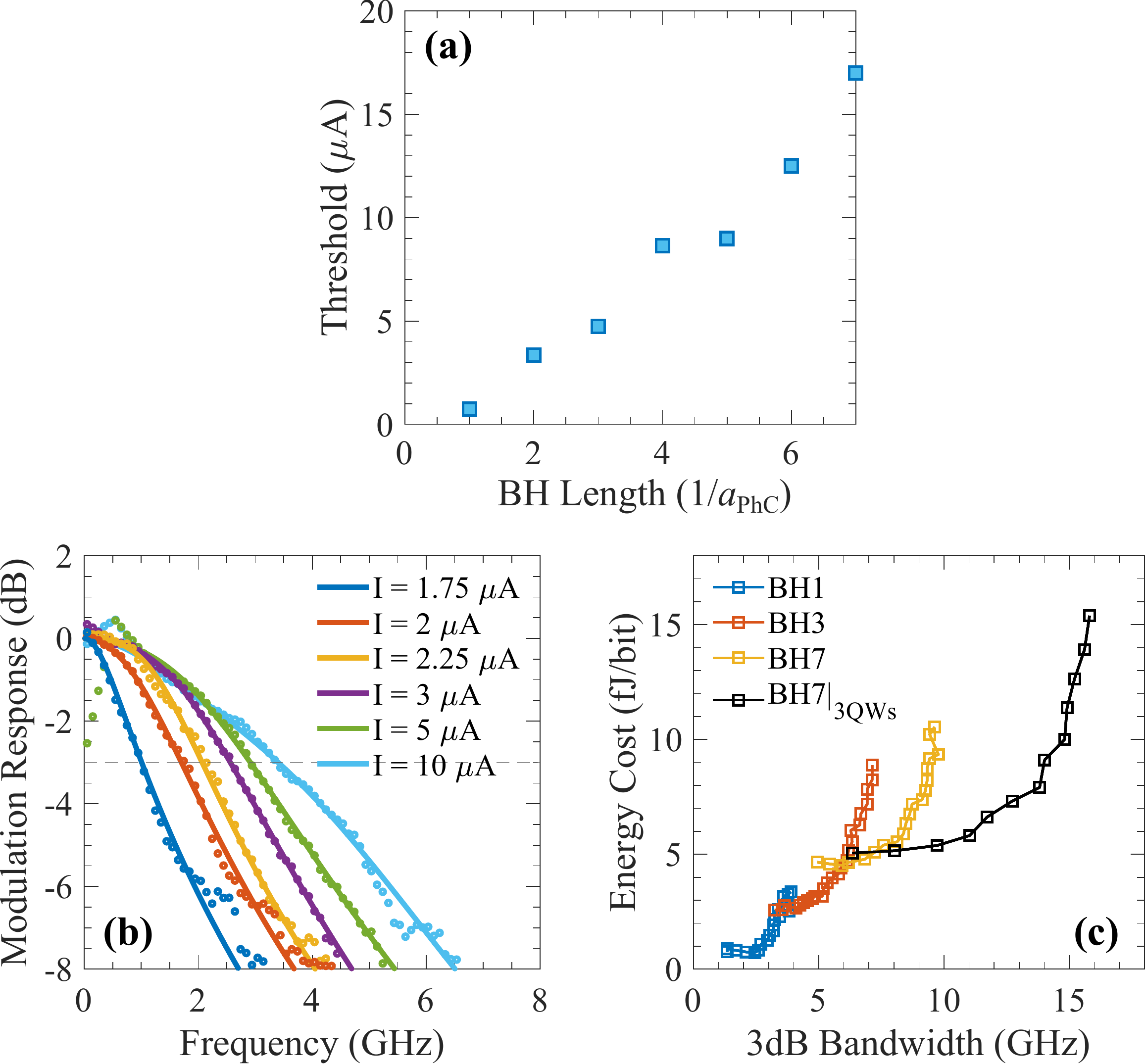}
    \caption{(a) Threshold current versus the BH length. (b) Small-signal response of the laser. (c) The energy cost per bit versus the modulation bandwidth for different active material sizes.}
    \label{fig:opt2}
\end{figure}

The laser threshold of L7 lasers with varying BH lengths is shown in Fig. \ref{fig:opt2}(a), exhibiting a linear-like reduction of the threshold current for shorter active regions. This confirms that the doping losses are the dominant loss channel, scaling similarly to the confinement factor. Theoretically, a characteristic BH length should exist below which the fixed mirror losses (undoped cavity) dominate, and the reduction of the modal gain leads to an increased threshold. However, such a characteristic length has not been observed in our experiments, suggesting that further miniaturization is possible.

In addition, we investigate the dynamic properties of the laser. Fig. \ref{fig:opt2}(b) shows the small-signal modulation response of the laser for different injected powers. The response function exhibits the characteristic bandwidth increase at higher injected currents. Due to the small cavity size and the high cavity Q-factor, the response is overdamped. The modulation efficiency of the laser is 38 GHz/mA\textsuperscript{0.5}, while the modulation frequency saturates above 3 GHz. In Fig. \ref{fig:opt2}(c), the energy cost per '1' bit for L7 cavity lasers with varying active material sizes is plotted against the 3-dB bandwidth. Each laser is identified by the length of its active region in units of the PhC lattice constant. In addition to the single QW design, a 3QW BH7 laser is shown for comparison. The energy cost is estimated as $EC = IV/1.3 f_{3dB}$ \cite{PhC3_Takeda2013}, where $f_{3dB}$ is the 3-dB bandwidth of the small-signal response function. The BH1 laser can operate at the typical processor clock rates of 2-3 GHz \cite{Miller2017} with less than 1 fJ/bit energy cost, which corresponds to an average below 0.5 fJ/bit for NRZ modulation with an equal number of '0' and '1' bits.

\section{Conclusions}
We have designed and experimentally demonstrated a photonic crystal nanolaser that exhibits an ultra-low threshold current of  730 nA in continuous-wave room-temperature operation. The laser is comprised of a sub-micron buried-heterostructure active region of a single quantum well and a matching doping profile that cause minimal absorption. Furthermore, small-signal modulation measurements showed a maximum 3-dB modulation bandwidth of 3 GHz that can be achieved with 1 fJ/bit energy cost. This laser type is the ideal candidate for future on-chip interconnects requiring efficient low-power consumption light sources.

\medskip
\textbf{Acknowledgements} \par 
Authors gratefully acknowledge funding by Villum Fonden via the NATEC Center of Excellence (Grant No. 8692), the European Research Council (ERC) under the European Union’s Horizon 2020 Research and Innovation Programme (Grant No. 834410 FANO), and the Danish National Research Foundation (Grant No. DNRF147 NanoPhoton).

\medskip
\textbf{Disclosures}\par
The authors declare no conflicts of interest.

\medskip
\textbf{Data Availability} \par
 The data that support the findings of this study are available from the corresponding author upon reasonable request.

\bibliography{bibtex.bib}

\begin{thebibliography}{11}%
\makeatletter
\providecommand \@ifxundefined [1]{%
 \@ifx{#1\undefined}
}%
\providecommand \@ifnum [1]{%
 \ifnum #1\expandafter \@firstoftwo
 \else \expandafter \@secondoftwo
 \fi
}%
\providecommand \@ifx [1]{%
 \ifx #1\expandafter \@firstoftwo
 \else \expandafter \@secondoftwo
 \fi
}%
\providecommand \natexlab [1]{#1}%
\providecommand \enquote  [1]{``#1''}%
\providecommand \bibnamefont  [1]{#1}%
\providecommand \bibfnamefont [1]{#1}%
\providecommand \citenamefont [1]{#1}%
\providecommand \href@noop [0]{\@secondoftwo}%
\providecommand \href [0]{\begingroup \@sanitize@url \@href}%
\providecommand \@href[1]{\@@startlink{#1}\@@href}%
\providecommand \@@href[1]{\endgroup#1\@@endlink}%
\providecommand \@sanitize@url [0]{\catcode `\\12\catcode `\$12\catcode
  `\&12\catcode `\#12\catcode `\^12\catcode `\_12\catcode `\%12\relax}%
\providecommand \@@startlink[1]{}%
\providecommand \@@endlink[0]{}%
\providecommand \url  [0]{\begingroup\@sanitize@url \@url }%
\providecommand \@url [1]{\endgroup\@href {#1}{\urlprefix }}%
\providecommand \urlprefix  [0]{URL }%
\providecommand \Eprint [0]{\href }%
\providecommand \doibase [0]{https://doi.org/}%
\providecommand \selectlanguage [0]{\@gobble}%
\providecommand \bibinfo  [0]{\@secondoftwo}%
\providecommand \bibfield  [0]{\@secondoftwo}%
\providecommand \translation [1]{[#1]}%
\providecommand \BibitemOpen [0]{}%
\providecommand \bibitemStop [0]{}%
\providecommand \bibitemNoStop [0]{.\EOS\space}%
\providecommand \EOS [0]{\spacefactor3000\relax}%
\providecommand \BibitemShut  [1]{\csname bibitem#1\endcsname}%
\let\auto@bib@innerbib\@empty
\bibitem [{\citenamefont {Miller}(2017)}]{Miller2017}%
  \BibitemOpen
  \bibfield  {author} {\bibinfo {author} {\bibfnamefont {D.~A.~B.}\
  \bibnamefont {Miller}},\ }\bibfield  {title} {\bibinfo {title} {Attojoule
  optoelectronics for low-energy information processing and communications},\
  }\href {https://doi.org/10.1109/JLT.2017.2647779} {\bibfield  {journal}
  {\bibinfo  {journal} {Journal of Lightwave Technology}\ }\textbf {\bibinfo
  {volume} {35}},\ \bibinfo {pages} {346} (\bibinfo {year} {2017})}\BibitemShut
  {NoStop}%
\bibitem [{\citenamefont {Takeda}\ \emph {et~al.}(2013)\citenamefont {Takeda},
  \citenamefont {Sato}, \citenamefont {Shinya}, \citenamefont {Nozaki},
  \citenamefont {Kobayashi}, \citenamefont {Taniyama}, \citenamefont {Notomi},
  \citenamefont {Hasebe}, \citenamefont {Kakitsuka},\ and\ \citenamefont
  {Matsuo}}]{PhC3_Takeda2013}%
  \BibitemOpen
  \bibfield  {author} {\bibinfo {author} {\bibfnamefont {K.}~\bibnamefont
  {Takeda}}, \bibinfo {author} {\bibfnamefont {T.}~\bibnamefont {Sato}},
  \bibinfo {author} {\bibfnamefont {A.}~\bibnamefont {Shinya}}, \bibinfo
  {author} {\bibfnamefont {K.}~\bibnamefont {Nozaki}}, \bibinfo {author}
  {\bibfnamefont {W.}~\bibnamefont {Kobayashi}}, \bibinfo {author}
  {\bibfnamefont {H.}~\bibnamefont {Taniyama}}, \bibinfo {author}
  {\bibfnamefont {M.}~\bibnamefont {Notomi}}, \bibinfo {author} {\bibfnamefont
  {K.}~\bibnamefont {Hasebe}}, \bibinfo {author} {\bibfnamefont
  {T.}~\bibnamefont {Kakitsuka}},\ and\ \bibinfo {author} {\bibfnamefont
  {S.}~\bibnamefont {Matsuo}},\ }\bibfield  {title} {\bibinfo {title}
  {{Few-fJ/bit data transmissions using directly modulated lambda-scale
  embedded active region photonic-crystal lasers}},\ }\href
  {https://doi.org/10.1038/nphoton.2013.110} {\bibfield  {journal} {\bibinfo
  {journal} {Nature Photonics}\ }\textbf {\bibinfo {volume} {7}},\ \bibinfo
  {pages} {569} (\bibinfo {year} {2013})}\BibitemShut {NoStop}%
\bibitem [{\citenamefont {Takeda}\ \emph {et~al.}(2015)\citenamefont {Takeda},
  \citenamefont {Sato}, \citenamefont {Fujii}, \citenamefont {Kuramochi},
  \citenamefont {Notomi}, \citenamefont {Hasebe}, \citenamefont {Kakitsuka},\
  and\ \citenamefont {Matsuo}}]{Takeda2014}%
  \BibitemOpen
  \bibfield  {author} {\bibinfo {author} {\bibfnamefont {K.}~\bibnamefont
  {Takeda}}, \bibinfo {author} {\bibfnamefont {T.}~\bibnamefont {Sato}},
  \bibinfo {author} {\bibfnamefont {T.}~\bibnamefont {Fujii}}, \bibinfo
  {author} {\bibfnamefont {E.}~\bibnamefont {Kuramochi}}, \bibinfo {author}
  {\bibfnamefont {M.}~\bibnamefont {Notomi}}, \bibinfo {author} {\bibfnamefont
  {K.}~\bibnamefont {Hasebe}}, \bibinfo {author} {\bibfnamefont
  {T.}~\bibnamefont {Kakitsuka}},\ and\ \bibinfo {author} {\bibfnamefont
  {S.}~\bibnamefont {Matsuo}},\ }\bibfield  {title} {\bibinfo {title}
  {{Heterogeneously integrated photonic-crystal lasers on silicon for on/off
  chip optical interconnects}},\ }\href {https://doi.org/10.1364/oe.23.000702}
  {\bibfield  {journal} {\bibinfo  {journal} {Optics Express}\ }\textbf
  {\bibinfo {volume} {23}},\ \bibinfo {pages} {702} (\bibinfo {year}
  {2015})}\BibitemShut {NoStop}%
\bibitem [{\citenamefont {Crosnier}\ \emph {et~al.}(2017)\citenamefont
  {Crosnier}, \citenamefont {Sanchez}, \citenamefont {Bouchoule}, \citenamefont
  {Monnier}, \citenamefont {Beaudoin}, \citenamefont {Sagnes}, \citenamefont
  {Raj},\ and\ \citenamefont {Raineri}}]{Crosnier2017}%
  \BibitemOpen
  \bibfield  {author} {\bibinfo {author} {\bibfnamefont {G.}~\bibnamefont
  {Crosnier}}, \bibinfo {author} {\bibfnamefont {D.}~\bibnamefont {Sanchez}},
  \bibinfo {author} {\bibfnamefont {S.}~\bibnamefont {Bouchoule}}, \bibinfo
  {author} {\bibfnamefont {P.}~\bibnamefont {Monnier}}, \bibinfo {author}
  {\bibfnamefont {G.}~\bibnamefont {Beaudoin}}, \bibinfo {author}
  {\bibfnamefont {I.}~\bibnamefont {Sagnes}}, \bibinfo {author} {\bibfnamefont
  {R.}~\bibnamefont {Raj}},\ and\ \bibinfo {author} {\bibfnamefont
  {F.}~\bibnamefont {Raineri}},\ }\bibfield  {title} {\bibinfo {title} {{Hybrid
  indium phosphide-on-silicon nanolaser diode}},\ }\href
  {https://doi.org/10.1038/nphoton.2017.56} {\bibfield  {journal} {\bibinfo
  {journal} {Nature Photonics}\ }\textbf {\bibinfo {volume} {11}},\ \bibinfo
  {pages} {297} (\bibinfo {year} {2017})}\BibitemShut {NoStop}%
\bibitem [{\citenamefont {Jeong}\ \emph {et~al.}(2013)\citenamefont {Jeong},
  \citenamefont {No}, \citenamefont {Hwang}, \citenamefont {Kim}, \citenamefont
  {Seo}, \citenamefont {Park},\ and\ \citenamefont {Lee}}]{PhC4_Jeong2013}%
  \BibitemOpen
  \bibfield  {author} {\bibinfo {author} {\bibfnamefont {K.-Y.}\ \bibnamefont
  {Jeong}}, \bibinfo {author} {\bibfnamefont {Y.-S.}\ \bibnamefont {No}},
  \bibinfo {author} {\bibfnamefont {Y.}~\bibnamefont {Hwang}}, \bibinfo
  {author} {\bibfnamefont {K.~S.}\ \bibnamefont {Kim}}, \bibinfo {author}
  {\bibfnamefont {M.-K.}\ \bibnamefont {Seo}}, \bibinfo {author} {\bibfnamefont
  {H.-G.}\ \bibnamefont {Park}},\ and\ \bibinfo {author} {\bibfnamefont
  {Y.-H.}\ \bibnamefont {Lee}},\ }\bibfield  {title} {\bibinfo {title}
  {Electrically driven nanobeam laser},\ }\href
  {https://doi.org/10.1038/ncomms3822} {\bibfield  {journal} {\bibinfo
  {journal} {Nature Communications}\ }\textbf {\bibinfo {volume} {4}},\
  \bibinfo {pages} {2822} (\bibinfo {year} {2013})}\BibitemShut {NoStop}%
\bibitem [{\citenamefont {Kuramochi}\ \emph {et~al.}(2018)\citenamefont
  {Kuramochi}, \citenamefont {Duprez}, \citenamefont {Kim}, \citenamefont
  {Takiguchi}, \citenamefont {Takeda}, \citenamefont {Fujii}, \citenamefont
  {Nozaki}, \citenamefont {Shinya}, \citenamefont {Sumikura}, \citenamefont
  {Taniyama}, \citenamefont {Matsuo},\ and\ \citenamefont
  {Notomi}}]{PhC5_Kuramochi2018}%
  \BibitemOpen
  \bibfield  {author} {\bibinfo {author} {\bibfnamefont {E.}~\bibnamefont
  {Kuramochi}}, \bibinfo {author} {\bibfnamefont {H.}~\bibnamefont {Duprez}},
  \bibinfo {author} {\bibfnamefont {J.}~\bibnamefont {Kim}}, \bibinfo {author}
  {\bibfnamefont {M.}~\bibnamefont {Takiguchi}}, \bibinfo {author}
  {\bibfnamefont {K.}~\bibnamefont {Takeda}}, \bibinfo {author} {\bibfnamefont
  {T.}~\bibnamefont {Fujii}}, \bibinfo {author} {\bibfnamefont
  {K.}~\bibnamefont {Nozaki}}, \bibinfo {author} {\bibfnamefont
  {A.}~\bibnamefont {Shinya}}, \bibinfo {author} {\bibfnamefont
  {H.}~\bibnamefont {Sumikura}}, \bibinfo {author} {\bibfnamefont
  {H.}~\bibnamefont {Taniyama}}, \bibinfo {author} {\bibfnamefont
  {S.}~\bibnamefont {Matsuo}},\ and\ \bibinfo {author} {\bibfnamefont
  {M.}~\bibnamefont {Notomi}},\ }\bibfield  {title} {\bibinfo {title} {{Room
  temperature continuous-wave nanolaser diode utilized by ultrahigh-Q few-cell
  photonic crystal nanocavities}},\ }\href
  {https://doi.org/10.1364/oe.26.026598} {\bibfield  {journal} {\bibinfo
  {journal} {Optics Express}\ }\textbf {\bibinfo {volume} {26}},\ \bibinfo
  {pages} {26598} (\bibinfo {year} {2018})}\BibitemShut {NoStop}%
\bibitem [{\citenamefont {Takeda}\ \emph {et~al.}(2021)\citenamefont {Takeda},
  \citenamefont {Tsurugaya}, \citenamefont {Fujii}, \citenamefont {Shinya},
  \citenamefont {Maeda}, \citenamefont {Tsuchizawa}, \citenamefont {Nishi},
  \citenamefont {Notomi}, \citenamefont {Kakitsuka},\ and\ \citenamefont
  {Matsuo}}]{Takeda2021}%
  \BibitemOpen
  \bibfield  {author} {\bibinfo {author} {\bibfnamefont {K.}~\bibnamefont
  {Takeda}}, \bibinfo {author} {\bibfnamefont {T.}~\bibnamefont {Tsurugaya}},
  \bibinfo {author} {\bibfnamefont {T.}~\bibnamefont {Fujii}}, \bibinfo
  {author} {\bibfnamefont {A.}~\bibnamefont {Shinya}}, \bibinfo {author}
  {\bibfnamefont {Y.}~\bibnamefont {Maeda}}, \bibinfo {author} {\bibfnamefont
  {T.}~\bibnamefont {Tsuchizawa}}, \bibinfo {author} {\bibfnamefont
  {H.}~\bibnamefont {Nishi}}, \bibinfo {author} {\bibfnamefont
  {M.}~\bibnamefont {Notomi}}, \bibinfo {author} {\bibfnamefont
  {T.}~\bibnamefont {Kakitsuka}},\ and\ \bibinfo {author} {\bibfnamefont
  {S.}~\bibnamefont {Matsuo}},\ }\bibfield  {title} {\bibinfo {title} {{Optical
  links on silicon photonic chips using ultralow-power consumption
  photonic-crystal lasers}},\ }\href {https://doi.org/10.1364/oe.427843}
  {\bibfield  {journal} {\bibinfo  {journal} {Optics Express}\ }\textbf
  {\bibinfo {volume} {29}},\ \bibinfo {pages} {26082} (\bibinfo {year}
  {2021})}\BibitemShut {NoStop}%
\bibitem [{\citenamefont {Dimopoulos}\ \emph {et~al.}(2022)\citenamefont
  {Dimopoulos}, \citenamefont {Sakanas}, \citenamefont {Marchevsky},
  \citenamefont {Xiong}, \citenamefont {Yu}, \citenamefont {Semenova},
  \citenamefont {M{\o}rk},\ and\ \citenamefont {Yvind}}]{Dimopoulos2022}%
  \BibitemOpen
  \bibfield  {author} {\bibinfo {author} {\bibfnamefont {E.}~\bibnamefont
  {Dimopoulos}}, \bibinfo {author} {\bibfnamefont {A.}~\bibnamefont {Sakanas}},
  \bibinfo {author} {\bibfnamefont {A.}~\bibnamefont {Marchevsky}}, \bibinfo
  {author} {\bibfnamefont {M.}~\bibnamefont {Xiong}}, \bibinfo {author}
  {\bibfnamefont {Y.}~\bibnamefont {Yu}}, \bibinfo {author} {\bibfnamefont
  {E.}~\bibnamefont {Semenova}}, \bibinfo {author} {\bibfnamefont
  {J.}~\bibnamefont {M{\o}rk}},\ and\ \bibinfo {author} {\bibfnamefont
  {K.}~\bibnamefont {Yvind}},\ }\bibfield  {title} {\bibinfo {title}
  {{Electrically-driven Photonic Crystal Lasers with Ultra-low Threshold}},\
  }\href {https://doi.org/10.1002/lpor.202200109} {\bibfield  {journal}
  {\bibinfo  {journal} {Laser \& Photonics Reviews}\ ,\ \bibinfo {pages}
  {2200109}} (\bibinfo {year} {2022})},\ \Eprint
  {https://arxiv.org/abs/2207.02931} {arXiv:2207.02931} \BibitemShut {NoStop}%
\bibitem [{\citenamefont {Coldren}\ and\ \citenamefont
  {Corzine}(1995)}]{Coldren1995DiodeLA}%
  \BibitemOpen
  \bibfield  {author} {\bibinfo {author} {\bibfnamefont {L.~A.}\ \bibnamefont
  {Coldren}}\ and\ \bibinfo {author} {\bibfnamefont {S.}~\bibnamefont
  {Corzine}},\ }\href@noop {} {\emph {\bibinfo {title} {Diode Lasers and
  Photonic Integrated Circuits}}}\ (\bibinfo  {publisher} {Wiley},\ \bibinfo
  {year} {1995})\BibitemShut {NoStop}%
\bibitem [{\citenamefont {Sakanas}\ \emph {et~al.}(2019)\citenamefont
  {Sakanas}, \citenamefont {Semenova}, \citenamefont {Ottaviano}, \citenamefont
  {Mørk},\ and\ \citenamefont {Yvind}}]{Sakanas2019}%
  \BibitemOpen
  \bibfield  {author} {\bibinfo {author} {\bibfnamefont {A.}~\bibnamefont
  {Sakanas}}, \bibinfo {author} {\bibfnamefont {E.}~\bibnamefont {Semenova}},
  \bibinfo {author} {\bibfnamefont {L.}~\bibnamefont {Ottaviano}}, \bibinfo
  {author} {\bibfnamefont {J.}~\bibnamefont {Mørk}},\ and\ \bibinfo {author}
  {\bibfnamefont {K.}~\bibnamefont {Yvind}},\ }\bibfield  {title} {\bibinfo
  {title} {Comparison of processing-induced deformations of inp bonded to si
  determined by e-beam metrology: Direct vs. adhesive bonding},\ }\href
  {https://doi.org/https://doi.org/10.1016/j.mee.2019.05.001} {\bibfield
  {journal} {\bibinfo  {journal} {Microelectronic Engineering}\ }\textbf
  {\bibinfo {volume} {214}},\ \bibinfo {pages} {93} (\bibinfo {year}
  {2019})}\BibitemShut {NoStop}%
\bibitem [{\citenamefont {Saldutti}\ \emph {et~al.}(2021)\citenamefont
  {Saldutti}, \citenamefont {Xiong}, \citenamefont {Dimopoulos}, \citenamefont
  {Yu}, \citenamefont {Gioannini},\ and\ \citenamefont
  {M{\o}rk}}]{Saldutti2021}%
  \BibitemOpen
  \bibfield  {author} {\bibinfo {author} {\bibfnamefont {M.}~\bibnamefont
  {Saldutti}}, \bibinfo {author} {\bibfnamefont {M.}~\bibnamefont {Xiong}},
  \bibinfo {author} {\bibfnamefont {E.}~\bibnamefont {Dimopoulos}}, \bibinfo
  {author} {\bibfnamefont {Y.}~\bibnamefont {Yu}}, \bibinfo {author}
  {\bibfnamefont {M.}~\bibnamefont {Gioannini}},\ and\ \bibinfo {author}
  {\bibfnamefont {J.}~\bibnamefont {M{\o}rk}},\ }\bibfield  {title} {\bibinfo
  {title} {{Modal Properties of Photonic Crystal Cavities and Applications to
  Lasers}},\ }\href {https://doi.org/10.3390/nano11113030} {\bibfield
  {journal} {\bibinfo  {journal} {Nanomaterials}\ }\textbf {\bibinfo {volume}
  {11}},\ \bibinfo {pages} {3030} (\bibinfo {year} {2021})}\BibitemShut
  {NoStop}%
\end{thebibliography}%

\end{document}